\documentclass[12pt]{article}
\usepackage{amsfonts}
\usepackage[margin=2cm,nohead]{geometry}
\usepackage{amsmath,latexsym,amssymb}
\usepackage{xcolor}
\usepackage[english]{babel}

\usepackage[normalem]{ulem}

\usepackage{xy}
\xyoption{matrix}
\xyoption{frame}
\xyoption{arrow}
\xyoption{arc}

\usepackage{ifpdf}
\ifpdf
\else
\PackageWarningNoLine{Qcircuit}{Qcircuit is loading in Postscript mode.  The Xy-pic options ps and dvips will be loaded.  If you wish to use other Postscript drivers for Xy-pic, you must modify the code in Qcircuit.tex}
\xyoption{ps}
\xyoption{dvips}
\fi

\entrymodifiers={!C\entrybox}

\newcommand{\ket}[1]{{\left\vert{#1}\right\rangle}}

\newtheorem{Theorem}{Theorem}
\newtheorem{Lemma}{Lemma}
\newtheorem{Comment}{Comment}
\newtheorem{Property}{Property}
\newtheorem{Definition}{Definition}

\newcommand{\Endproof}{\hfill$\Box$ \\}

\begin{document}

\title{Quantum search in a dictionary based on fingerprinting-hashing}

\author{Farid Ablayev  \and  Nailya Salikhova
\and  Marat Ablayev}

\date{}


\maketitle

\begin{abstract}
In this work, we present a quantum query algorithm for searching a word of length $m$ in an unsorted dictionary of size $n$. The algorithm uses $O(\sqrt{n})$ queries (Grover operators), like previously known algorithms.

What is new is that the algorithm is based on the quantum fingerprinting-hashing technique, which (a) provides a first level of amplitude amplification before applying the sequence of Grover amplitude amplification operators and (b) makes the algorithm more efficient in terms of memory use --- it requires $O(\log n + \log m)$ qubits.
Note that previously developed algorithms by other researchers without hashing require $O(\log n + m)$ qubits.

\end{abstract}
\section{Introduction}

The problems of information retrieval in databases are well known in theoretical and applied computer science. The problem is an  enumeration problem, which classically requires an  enumeration of all possible options. The well-known classical unsorted list search is a sequential search query algorithm. Such algorithms, for our convenience, can be  described as follows.
We need to solve a Boolean equation $f(x)=1$, where $f$ is a Boolean function on  $x\in V$, $V= \{ w_0, \dots , w_{n-1} : w_i\in  \{0,1\}^m\}$. It is assumed that the function  $f$ is given as  an oracle. In the query model, one can ask the oracle only a question like: ``what is $f$ given $x$?'', and use the answer in further computations. The number of queries is a characteristic of the complexity (time) of such a query algorithm.

The above problem of solving the equation $f(x) = 1$ is a general form of the iteration problem, which classically requires successive iteration over all elements of $V$ for the case where $V$ is unsorted. The quantum algorithm finds some root of the equation using $\frac{\pi}{4}\sqrt{|V|}$ calls to the function $f$. 

Such a quantum search, which accelerates search by a quadratic factor compared to classical algorithms, was proposed by L. Grover in 1996 \cite{g96} and has been modified in a number of subsequent works (see, the paper \cite{brassard2002quantum}, lecture notes \cite{de2019quantum}  and  the book  \cite{kaye2006introduction} for more information and citations). The development of quantum algorithms for information retrieval in databases continues. For example, see reviews of \cite{DBLP:journals/bigdatama/AblayevAHKSW20, DBLP:journals/bigdatama/AblayevAHKSW20a}.

\paragraph{Search for a word in the text.} Searching for a word in a text, as a task of searching for information in a database, has been considered in many works and has some specific features. First, the oracle $f_w$, defined by the task, is used to find the occurrence of the word $w$ in the dictionary $V$. Namely, $f_w(x)= 1$ if and only if $w=x$. Secondly, for searching for a word in a text to speed up search in the 1970s and 1980s, hashing-based algorithms were proposed. The Knuth-Morris-Pratt (KMP) algorithm \cite{knuth1977fast} and the Rabin-Karp algorithm \cite{karp1987efficient} solve the problem in linear number of comparisons (in linear time), with $O(n+m)$ complexity.

It is clear that quantum algorithms for this task, based on Grover's idea, provide a quadratic saving in the number of calls to the oracle. Note that to describe Grover's algorithm, we first need to specify an efficient  oracle circuit for  $f_w$ for such  algorithm. A number of similar studies in recent decades in the field of developing algorithms for searching for occurrences of a word $w$ in a text $V$ (in an unsorted dictionary $V$ constructed from the text) were devoted to the construction of efficient oracles $f_w$. A number of constructions of such oracles in terms of quantum circuits were proposed and the complexity of such circuits was studied: Practical Implementation of a Quantum String Matching
Algorithm \cite{marino2024practical}, Algorithm of Ramesh, Hariharan, and Vinay \cite{ramesh2003string} (2003). Montanaro \cite{montanaro2017quantum} (2017). Soni and Rasool \cite{soni2020pattern} (2020).

Note that all of these listed results require $O(\log n + m )$ qubits to search for a word $w$ of length $m$ in a dictionary $V$ of size $n$.

The idea of using hashing methods for quantum information retrieval in an unordered database was realized in 2024 \cite{ablayev2024hybrid}. In \cite{ablayev2024hybrid}, a hybrid classical-quantum (probabilistic-quantum) algorithm is proposed for searching a word $w$ of length $m$ in a dictionary $V$ of size $n$.

The key idea of the hybrid classical-quantum algorithm is as follows: (a) choose a universal hash family $\cal F$ of hash functions, (b) uniformly randomly choose a hash function $h\in {\cal F}$, (c) hash the elements of $V$ with the hash function $h$, and (d) apply the quantum amplification technique to search for the hash image $h(w)$ of the word $w$ in the hashed dictionary $h(V)$.
It has been proven that using hashing methods can exponentially reduce the number of qubits required. Specifically, $ \log n + \log m$ qubits are sufficient when using hashing, instead of $\log n + m$ qubits in algorithms without hashing.

\paragraph{ Quantum fingerprinting-hashing.} 
The quantum fingerprinting-hashing technique (quan\-tum fingerprinting-hashing technique or quantum hashing technique for short) is a key component of the algorithms considered in this paper.
The quantum fingerprinting   was formalized as a method of quantum information compression in 2001 by Burman et al. \cite{buhrman2001quantum}. The quantum fingerprinting function presented in \cite{buhrman2001quantum} was built on error-correcting codes. Various designs of quantum hash functions and their cryptographic properties were discussed in \cite{DBLP:journals/bjmc/AblayevAVZ16}. See also  the book \cite{aav_book_2023}.  The quantum hash function $\psi : \Sigma^m\to ({\mathcal H}^2)^{\otimes s}$ maps words of length $m$ to $s$ qubit states. Such a map has the following important (for our algorithms) properties: (a) the map is contractive, i.e. $s<m$ (for example for fingerprinting function \cite{buhrman2001quantum} $s=\log m$), (b) the resulting quantum states are highly distinguishable, and (c) the function $\psi$ is invertible.  
\paragraph{Our contribution.} In this work, we 
make a further step in applying hashing technique for  finding a specific word of length $m$ from an unordered collection (vocabulary) of $n$ words, each of which has length $m$. 

The quantum search algorithm works as follows. The vocabulary $V$ is represented as a quantum state $\ket{V,\psi}$, whose elements are hashed by the quantum hash function $\psi$. The search for an occurrence of the word $w$ is performed in  two stages as follows. In the first stage, the quantum state is quantum-parallel inverted by the mapping $\psi^{-1}(w)$ determined by $w$. This stage provides the first level of amplitude amplification. Then, in  the second stage, a known sequence of Grover amplitude amplification operators is applied. Finally, the resulting quantum state is measured, and the algorithm produces the result. 

The high probability of the correct result is based on properties (b) and (c) of the quantum hash function --- they provides a first level of amplitude amplification before applying the sequence of Grover amplitude amplification operators

The quantum hash function is memory efficient: searching for a word $w$ of length $m$ in a dictionary $V$ of size $n$ requires $O(\log n + s )$ qubits. The number of oracle calls is $O(\sqrt{n})$.

Note that, as in the previous work \cite{ablayev2024hybrid}, we present the algorithm in terms of the quantum query model. This level of representation is sufficient to demonstrate the efficiency of memory use by the algorithm. We leave the issues of efficient implementation of the oracle scheme for further consideration.\\

The paper is organized as follows. In Section 2, we describe the quantum fingerprinting function and the necessary operators for its further implementation. In Section 3, we present a dictionary search algorithm based on the quantum fingerprinting function in technical detail. In Section 4, we formulate and prove theorems analyzing the operation of the algorithm and its complexity characteristics.

In Section 5.2, we generalize the specific construction of the algorithm from Section 3 to the general case of constructing ${\cal A}$ algorithms based on the quantum hashing technique in terms of an arbitrary $\epsilon$-collision-resistant quantum hash function $\psi$.  More precisely, in Section 5 we present a group of algorithms that can differ depending on which quantum hash function (see Section 5.1) we use. 

In Section 6, we present its characteristics. We conclude Section 6 with a discussion of the features of the algorithm.

\section{Quantum fingerprinting-hashing}
The quantum function $\psi$, defined in this section, is considered in the paper \cite{Buhrman:2001:Fingerprinting}. The authors called their function ``quantum fingerprinting function''.  The $\psi$ function is defined based on a binary error-correcting code. 
\paragraph*{Error Correcting Codes.} 
{\em The error-correcting code} is defined by the mapping
\[ E : \Sigma^m \to \Sigma^l \]
with the following condition:
for any different words
\[
w,w'\in\Sigma^m,
\]

their maps 
\[
E(w),E(w')\in \Sigma^l
\] 
satisfy the condition that the Hamming distance  $d(E(w),E(w'))$ between them is at least $d$. 

Such a code $E$ is called an $(l,m,d)$-code. A binary code is one in which $\Sigma =\{0,1\}$. 

{\em For arbitrary $d>1$, there exists an $(l,m,d)$-error-correcting code $E$ with $l=cm$, $c>1$.}

\paragraph*{Function $\psi_E$ generated by  error correcting code.} 
The function $\psi_E$ generated by the binary $(l,m,d)$-error-correcting code $E$ defined as follows. 

For $s=\log l$ the function  
\[ 
 \psi_{E}  :  \{0,1\}^m\to({\cal H}^2)^{\otimes {(s+1)}}    
\]
is defined by the condition
 \begin{equation}\label{psi_E}
  \ket{\psi_{E}(w)} = \frac{1}{\sqrt{2^s}}\sum_{i=0}^{2^s-1}\ket{i} \ket{E_i(w)},
\end{equation}
where $E_i(w)$ is the $i$-th bit of the codeword $E(w)$. 

In content: a word $w$ with length $m$ is mapped to a $(\log m +const)$-qubit quantum state $\ket{\psi_{E}(w)}$. This state $\ket{\psi_{E}(w)}$ represents the encoding $E(w)$ of the original word $w$.  The state $\ket{\psi_{E}(w)}$ is a superposition over $s+1 =\log m +const$ qubits. 

\paragraph*{Realization of the quantum fingerprinting function $\psi_E$.} 
Let $\ket{\bf 0}=\ket{0}^{\otimes {(s+1)}}$.

 The transformation 
\begin{equation}\label{U1}
 U_{\psi_E} : \ket{\bf 0} \xrightarrow{w} ({\mathcal H}^2)^{\otimes(s+1)}
 \end{equation}
 acts on $(s+1)$-qubits. 
 We define the transformation $U_{\psi_E}(w) =U_E(w) (H^{\otimes s} \otimes I)$. It is defined by a $2^{s+1}\times2^{s+1}$ unitary matrix. It is a composition of the Hadamard transformation $H^{\otimes s}$, the identity transformation $I$ and the transformation $U_E$. It is convenient to define the $U_E$ transformation by describing its action on the $2^{s+1}$ vectors $\ket{k}$ of the computational basis $B=\{(0,\dots, 1), \dots, (1,\dots, 0)\}$ of the $({\mathcal H}^2)^{\otimes(s+1)}$ space. The arguments of the transformation $U_E$ are words $w\in\{0,1\}^m$: 
\[ {U_E} : \ket{k}\xrightarrow{w} \ket{k'}. 
\]
${U_E}$ is defined by the codeword $E(w)$, which has a length of $l=2^s$.  In terms of content, the transformation $U_E$ ``changes'' the last $(s+1)$-st qubit $\ket{a}$ in the basis state $\ket{k}=\ket{i}\otimes\ket{a}$ to $\ket{k'}=\ket{i}\otimes\ket{a\oplus E_i(w)}$. 
\begin{Property}\label{u_property}
Transformation 
  $U_{\psi_E}(w) ={U_E}(w) (H^{\otimes s} \otimes I)$ defines the function $\psi_E$. Namely, for any word $w\in\{0,1\}^m$ the following is true
\[
 \ket{\psi_{E}(w)} = U_{\psi_E}(w)\ket{\bf 0}.
\]
\end{Property}
{\em Proof.}
The transformation $H^{\otimes s} \otimes I$, transforms the basis state $\ket{0}^{\otimes {(s+1)}}$ into an equal-amplitude superposition of the first $s$ qubits (Hadamard transformation $H^{\otimes s}$ of the first $s$ qubits)
\[ \ket{\bf 0}=\ket{0}^{\otimes {(s+1)}} \xrightarrow{H^{\otimes s} \otimes I}
\ket{\psi_0}=\frac{1}{\sqrt{2^s}}\sum\limits_{i=0}^{2^s-1}\ket{i}\otimes \ket{0}.
\]
Next, the transformation ${U_E}(w)$ transforms the state $\ket{\psi_0}$ to the state $\ket{\psi_E(w)}$:
\[
\frac{1}{\sqrt{2^s}}\sum\limits_{i=0}^{2^s-1}\ket{i}\otimes \ket{0}   \xrightarrow{{U_E}(w)} \ket{\psi_E(w)}=\frac{1}{\sqrt{2^s}}\sum_{i=0}^{2^s-1}\ket{i} \ket{0\oplus E_i(w)}= \frac{1}{\sqrt{2^s}}\sum_{i=0}^{2^s-1}\ket{i} \ket{E_i(w)}. 
\]
\Endproof
%
 Next, we also need the transformation 
  \[
 U^{-1}_{\psi_E} : ({\mathcal H}^2)^{\otimes(s+1)} \xrightarrow{w}  ({\mathcal H}^2)^{\otimes(s+1)},
 \]
 given by words $w\in\{0,1\}^m$, and it is the inverse of the transformation $U_{\psi_E}$ in the following sense
 \[ 
 U^{-1}_{\psi_E} : \ket{\psi_E(w)} \xrightarrow{w}  \ket{\bf 0}. 
 \] 
 
\begin{Property}\label{u-1_property}
The transformation $U^{-1}_{\psi_E}(w)= (H^{\otimes s} \otimes I) \otimes U_E(w)$ is the inverse of the transformation $U_{\psi_E}(w)$.  
\end{Property}
 {\em Proof. }
  Indeed, for an arbitrary word $w\in\{0,1\}^m$, for a $(s+1)$-qubit state $\ket{\psi_E(w)}$ it is true that $U^{-1}_{\psi_E}(w)\ket{\psi_E(w)}= \ket{\bf 0}$:
\[
\frac{1}{\sqrt{2^s}}\sum_{i=0}^{2^s-1}\ket{i} \ket{E_i(w)} \xrightarrow{U_E(w)} \frac{1}{\sqrt{2^s}}\sum_{i=0}^{2^s-1}\ket{i} \ket{E_i(w)\oplus E_i(w)}=  \frac{1}{\sqrt{2^s}}\sum\limits_{i=0}^{2^s-1}\ket{i}\otimes \ket{0} \xrightarrow{H^{\otimes s} \otimes I} \ket{0}^{\otimes {s+1}}. 
\]
\Endproof

 \section{Algorithm  ${\cal A}$. }

 \paragraph{Problem}

Given an unordered set $V$ composed of $n$ binary sequences $w_k$, each of length $m$. 
\[V=\{w_0,\dots , w_{n-1} \},\]
where $w_k=\{0,1\}^m$ for $0\le k \le n-1$.

Given a binary string $w$ of length $m$, where $m<n$. 
It is required to find the index of the occurrence of $w$ in the sequence $V$. Namely, it is required to find an index $k$ such that $w=w_{k}$.

Note that this is essentially a database search problem.

\paragraph{Algorithm}

 The algorithm ${\cal A}$ consists of two parts:
\begin{enumerate}
\item First part: preparing the initial state based on the sequence $V$.
\item Second part: reading the search word $w$ and searching for its position in the sequence.
\end{enumerate}

 Note that the algorithm  ${\cal A}$ takes different input data in the two parts: $V$ and $w$, respectively.

{\bf Description of the algorithm ${\cal A}$.}
\begin{quote}
\item {\bf Input data:}

{\bf For the first part}: the sequence $V=\{w_0, \dots, w_{n-1}\}$ of binary words with a length of $m$.

{\bf For the second part}: a binary string $w$ with length $m$.

\item {\bf Output data:} The index $k$, which denotes the number of an element in the sequence, such that $w=w_k$.
\end{quote}

Thus, the algorithm ${\cal{A}}$ implements the mapping 
\[ {\cal{A}}: V, w \longmapsto k. \]

\begin{enumerate} 

\item {\bf The first part of the algorithm (Initialization of hashed vocabulary)} involves preparing the initial state $\ket{V, \psi_E}$.

The initial state is 
\[
\ket{\psi_0} =  \frac{1}{\sqrt{n}}\sum\limits_{j=0}^{n-1}\ket{j} \otimes \ket{\bf 0} \otimes \ket{1}.
\]
Here, $\ket{\bf 0}$ -- is the zero state of $s+1$ qubits.
According to this sequence $V$, based on the transformation $U_{\psi_E}$, which defines the quantum function $\psi_E$, the state  $\ket{0}$ is converted into the initial  state:
    \[
\ket{V, \psi_E} = \frac{1}{\sqrt{n}}\sum\limits_{j=0}^{n-1}\ket{j} \otimes \ket{\psi_E(w_j)} \otimes \ket{1}.
\] 
Recall that 
\begin{equation}\label{psi_E}
  \ket{\psi_{E}(w_j)}  = \frac{1}{\sqrt{2^s}}\sum_{i=0}^{2^s-1}\ket{i} \ket{E_i(w_j)},
\end{equation}

\item {\bf The second part of the algorithm} ${\cal{A}}$ consists of reading the searched word $w$ and searching for a position $k$, such that $w=w_k$.

The search procedure consists of applying the amplitude amplification method presented in the paper \cite{brassard2002quantum} and described in the book \cite{kaye2006introduction}.

\begin{itemize}
    \item \textbf{Conversion hash transformation} 
    
    The $I^{\otimes \log{n}} \otimes U_{\psi_E}^{-1}(w) \otimes I$ operator is applied to the state $\ket{V, \psi_E}$,
\begin{eqnarray}
  \ket{V, \psi_E, w} & = &
\frac{1}{\sqrt{n}}\sum\limits_{j=0}^{n-1}\ket{j}  \left (\frac{1}{\sqrt{2^s}}\sum\limits_{i=0}^{2^s-1}\ket{i}\ket{E_i(\omega_j) \oplus E_i(\omega)} \right)  \ket{1} \nonumber
\\ 
 &= & \frac{1}{\sqrt{n}}\sum\limits_{j=0}^{n-1}\ket{j} \left ( \frac{1}{\sqrt{2^s}}\sum\limits_{i=0}^{2^s-1}\ket{i} \left (\sum\limits_{i:E_i(\omega_j)=E_i(\omega)}\ket{0} + \sum\limits_{i:E_i(\omega_j) \neq E_i(\omega)}\ket{1}\right) \right) \ket{1}  \nonumber 
 \\
&=&  \frac{1}{\sqrt{n}} \sum\limits_{j=0}^{n-1}\ket{j}  \ket{\phi_E(w_j, w)} \ket{1}, \nonumber
\end{eqnarray}

where $ \ket{\phi_E(w_j, w)}= U^{-1}_{\psi_E}(w)\ket{\psi_E(w_j)}$ for $j \in\{0,\dots, n-1\}$. 

\item \textbf{Amplification} 

The amplification procedure of the amplitudes of basic states $\ket{j}\ket{0}^{\otimes s+1}\ket{1}$ (``good states'' as they are called in \cite{kaye2006introduction}) is applied to state $\ket{V, \psi_E, w}$.

The amplitude amplification procedure consists in the use of Grover macro steps according to the description in \cite{kaye2006introduction}.

\item \textbf{Measurement} 

The first $\log{n} +s+1$ qubits of the final state are measured in a computational basis. If the last $s+1$ qubits are all zero, then the result of measuring the first $\log n$ qubits is the position of element $w_k$, where $w_k=w$.
\end{itemize}




\end{enumerate}

\subsection{ Characteristics of the ${\cal A}$ algorithm.} 

The characteristics of a quantum algorithm include: the probability of error, the number of queries made to the analyzed data, and the amount of memory used.

{\em The probability of success.}
We denote by $Pr_{success}({\cal A})$  the probability of a successful outcome of the algorithm ${\cal A}$.

{\em Query complexity.} The number of queries $Q({\cal A})$ (the number of requests to the oracle) is the query complexity of the quantum algorithm ${\cal A}$.

{\em Memory complexity.} The number $S({\cal A})$ of used qubits is a measure of the memory complexity of quantum algorithm ${\cal A}$.

\section{Analysis of the ${\cal A}$ algorithm.} 

In this section, we define the theorem \ref{th_aa_formal} and present its proof. The main states result is that the algorithm ${\cal A}$  provides the correct result quickly with high probability and low memory complexity. Informally, this result is presented in theorem \ref{th_aa_informal}.

Denote by $Pr_{success}({\cal A})$ the probability of the event that the result of algorithm ${\cal A}$ is a number $k$, such that $w_k=w$.

Meaningfully, the algorithm $\cal A$ has the following characteristics.

\begin{Theorem} \label{th_aa_informal}   
For the algorithm ${\cal A}$ which searches for an occurrence of an element with length $m$ in a sequence of $n$ elements, the following is true
 \[ Pr_{success}({\cal A}) \approx 1, \quad Q({\cal A}) = O(\sqrt{n}) \quad \mbox{ and } \quad S({\cal A})  = O(\log n +\log m).
\]
 
\end{Theorem}

The   Theorem \ref{th_aa_informal}  is based on the  following formal statement.

\begin{Theorem}\label{th_aa_formal}  
Let  $c>0$,  $t = O(\sqrt{n})$ and $a=\sin^2((2t+1)\theta)$ , where $sin(\theta)\in\left(\sqrt{1/n}, \sqrt{1/n+ c}\right] $.  

 \[ Pr_{success}({\cal A}) \ge a\frac{1}{1+c}, \quad Q({\cal A}) = O(\sqrt{n}) 
\quad \mbox{ and  } \quad S({\cal A}) \le 2 \log n +\log m + const.
\]

 \end{Theorem}

We present the proof of Theorem  \ref{th_aa_formal}  in the next section.  We conclude that section  by demonstrating  that parameters $a$ and $c$ can be selected to make  $Pr_{success}({\cal A})$ close to 1. 

\subsection{Proof of the theorem \ref{th_aa_formal}.}

In this section, we present a proof of the estimates for Theorem \ref{th_aa_formal}. For convenience, the evaluation of each characteristic has been presented in its own section.

\subsubsection{Estimation of the probability $Pr_{success} ({\cal A})$ of the algorithm success and query complexity $Q({\cal A})$}

The following Property \ref{buhtman2001} is key to the assertion of the Lemma \ref{m_lemma}. This Lemma \ref{m_lemma}, in turn, is key to proving the high probability of correct operation of the algorithm.

\begin{Property}\label{buhtman2001}
    Let  the function $\psi_E$ (\ref{psi_E}) generated by a binary  error correcting $(l,m,d)$-code $E$  with $d\ge(1-\epsilon)l$ for some  $\epsilon \in(0,1)$.
Then for an arbitrary pair of different words $w,w'\in\{0,1\}^m$ the following is true
\[
\left|\langle{\psi_E(w)}|{\psi_E(w')}\rangle\right| \le
\epsilon. 
\]
\end{Property} 
{\em Proof. }See \cite{Buhrman:2001:Fingerprinting} and the book \cite{aav_book_2023} for the proof.\Endproof

In this case, it is natural to say that the states $\ket{\psi_E(w)}$ and $\ket{\psi_E(w')}$ are $\epsilon$-orthogonal.

\begin{Lemma}\label{m_lemma}
 Let  the function $\psi_E$ (\ref{psi_E}) generated by a binary  error correcting $(l,m,d)$-code $E$  with $d\ge(1-\epsilon)l$ for some  $\epsilon \in(0,1)$. Then for the state $ \ket{\phi_E(w_j, w)}= U^{-1}_{\psi_E}(w)\ket{\psi_E(w_j)}$ 
 \[\ket{\phi_E(w_j,w )}=\alpha_0\ket{0}+ \alpha_1\ket{1} + \dots +  \alpha_{2^l-1}\ket{2^s-1}  \]
the following is true. 
If  $w_j=w$, then  
$ \alpha_0=1 
$.
If $w_j\not= w$, then  
   $ \alpha_0 \le \epsilon  
   $.
\end{Lemma}
 {\em Proof.} 
%
The transformation $U^{-1}_E(w)$ is the inverse of $U_E(w)$.  $U^{-1}_E(w)$ is applied  to  state $\ket{\psi_E(w_j)}$ to get  $\ket{\phi_E(w_j,w )}$.  Therefore, for the word $w_k=w$
\[\ket{\phi_E(w_k,w)}=\ket{0}= 1\ket{0}+0\ket{1}+\dots +0\ket{2^s-1}. \]
For all other words $w_j\in\{0,1\}^m$, their function values $\ket{\psi_E(w_k)}$ and $\ket{\psi_E(w_j)}$ are pairwise $\epsilon$-orthogonal (due to the Property \ref{buhtman2001}).
\[
\left|\langle{\psi_E(w_k)}|{\psi_E(w_j)}\rangle\right| \le
\epsilon. 
\]

Unitary transformation  $U^{-1}_E(w)$ of states  $\ket{\psi_E(w_k)}$ and $\ket{\psi_E(w_j)}$ saves the scalar product. This means that all states $\ket{\phi_E(w_j,w)} = U^{-1}_E(w)\ket{\psi_E(w_j})$ for $w_j\not= w$,  $w_j\in \{0,1\}^m$ are pairwise $\epsilon$-orthogonal
\[
\left|\langle{\phi_E(w_k,w)}|{\phi_E(w_j,w)}\rangle\right| \le
\epsilon. 
\]
Therefore, for states
\[\ket{\phi_E(w_j,w )}=\alpha_0\ket{0}+ \alpha_1\ket{1} + \dots +  \alpha_{2^l-1}\ket{2^s-1}  \]
for  $w_j\not= w$ it is true that $|\alpha_0|= \epsilon_j \le \epsilon$. 
The latter proves the statement of the lemma. \Endproof

\paragraph*{Amplification.} The amplification procedure corresponds to the one presented in the article \cite{brassard2002quantum}. For convenience, we will follow the description and notations of the procedure in chapter 8 of the book \cite{kaye2006introduction}.

The amplification procedure is implemented by repeatedly applying the unitary operator $QQ$ to the state $\ket{V, \psi_E, w}$. In the text below, for simplicity, we will use $\ket{\psi}$ to indicate the state of $\ket{V, \psi_E, w}$.

To prove the theorem, we first divide the vector $\ket{\psi}$ into two parts $\ket{\psi_0}$ and $\ket{\psi_1}$.

\[
\ket{\psi}= \ket{\psi_0} + \ket{\psi_1}, 
\]
which, according to the Lemma \ref{m_lemma}
\begin{equation}\label{psi_0}
   \ket{\psi_0}= \frac{1}{\sqrt{n}}\left(\ket{k}\ket{0}\ket{1} + \sum_{j=0, j\not= k}^{n-1}\epsilon_j\ket{j}\ket{0}\ket{1} \right)
\end{equation}

and the vector $\ket{\psi_1}$ consists of the remaining components of the state $\ket{\psi}$. Let's denote
\[
p_{good}= \frac{1}{n}\left( 1 + \sum_{j=0, j\not= k}^{n-1}\epsilon_j^2\right) \quad \mbox{and } \quad p_{bad}=1-p_{good}.
\]

$p_{good}$ is the probability of measuring $n$ basic states $\ket{j}\ket{0}\ket{1}$ for $j\in\{0, \dots n-1\}$. We will call these basic states good states, and the probability of obtaining them good probability $p_{good}$. We will call other basic states $\ket{j}\ket{i}\ket{1}$ bad states, with the probability of obtaining them being called bad probability $p_{bad}$. We renormalize components $\ket{\psi_0}$ and $\ket{\psi_1}$ and get the following states 
\begin{equation}\label{psi_good}
 \ket{\psi_{good}}= \frac{1}{\sqrt{p_{good}}}\ket{\psi_0} \quad  \mbox{and } \quad \ket{\psi_{bad}}= \frac{1}{\sqrt{p_{bad}}}\ket{\psi_1}.   
\end{equation}

Then we can record
\[
\ket{\psi}= \sqrt{p_{good}}\ket{\psi_{good}} + \sqrt{p_{bad}}\ket{\psi_{bad}}, 
\]
or 
\[
\ket{\psi}= \sin(\theta)\ket{\psi_{good}} + \cos(\theta)\ket{\psi_{bad}}, 
\]
where $\theta\in\left(0, \frac{\pi}{2}\right)$ defined by equality $\sin^2(\theta)=p_{good}$. \\

We define the search iteration operator $QQ = U_{\psi}^{\perp}U_{f}$ as follows.

For an arbitrary real number $\theta$, the operation $U_{f}$ acts as follows:
\[
U_{f}(\sin(\theta)\ket{\psi_{good}} + \cos(\theta)\ket{\psi_{bad}}) = -\sin(\theta)\ket{\psi_{good}} + \cos(\theta)\ket{\psi_{bad}}
\]
and thus, $U_{f}$ performs reflection relative to the axis defined by the vector $\ket{\psi_{bad}}$.

More precisely, the operator $U_{f}$ changes the amplitude sign for the basic states of the form $\ket{j}\ket{0}\ket{1}$.

Let's denote the state
\[
\ket{\overline{\psi}} = \cos(\theta)\ket{\psi_{good}} - \sin(\theta)\ket{\psi_{bad}}
\]
which is orthogonal to $\ket{\psi}$.
\[
\{\ket{\psi_{good}}, \ket{\psi_{bad}}\}
\]
and
\[
\{\ket{\overline{\psi}}, \ket{\psi}\}
\]
are orthonormal bases for the same 2-dimensional space.
\[
U_{f}\ket{\psi} = -\sin(\theta)\ket{\psi_{good}} + \cos(\theta)\ket{\psi_{bad}} = \cos(2\theta)\ket{\psi} - \sin(2\theta)\ket{\overline{\psi}}
\]

The operator $U_{\psi}^{\perp}$ acts as follows:
\[
U_{\psi}^{\perp}(\sin(\theta)\ket{\psi} + \cos(\theta)\ket{\overline{\psi}}) = \sin(\theta)\ket{\psi} - \cos(\theta)\ket{\overline{\psi}}
\]
and thus, $U_{\psi}^{\perp}$ performs reflection relative to an axis defined by a vector $\ket{\psi}$.\\

Indeed, 
\[
U_{\psi}^{\perp}U_{f}\ket{\psi} = U_{\psi}^{\perp}(-\sin(\theta)\ket{\psi_{good}} + \cos(\theta)\ket{\psi_{bad}}) = \cos(2\theta)\ket{\psi} + \sin(2\theta)\ket{\overline{\psi}}
\]
and can be described in the basis
\[
\{\ket{\psi_{good}}, \ket{\psi_{bad}}\}
\]
as 
\[
U_{\psi}^{\perp}U_{f}\ket{\psi} = \sin(3\theta)\ket{\psi_{good}} + \cos(3\theta)\ket{\psi_{bad}}
\]

Repeating the application of the operator $QQ$ $t$ once brings the initial state $\ket{\psi}$ to
\begin{equation}\label{psi_t}
  \ket{\psi^t}= QQ^t\ket{\psi}=  \sin((2t+1)\theta)\ket{\psi_{good}} +\cos((2t+1)\theta)\ket{\psi_{bad}}   
\end{equation}

Let $a=\sin^2((2t+1)\theta)$. Note that for small $\theta$: $\sin(\theta)\ge\theta-\delta$ for small $\delta$ ($\sin(\theta)\approx \theta$). In particular, in our case
\[
\sin(\theta)= \sqrt{\frac{1}{n}\left( 1 + \sum_{j=0, j\not= k}^{n-1}\epsilon_j^2\right)}
\]
and $sin(\theta)\in\left(\sqrt{1/n}, \sqrt{1/n+ c}\right] $, for small $c>0$. We choose $t$ such that $(2t+1)\theta \approx\frac{\pi}{2}$, meaning $t\in\Omega(\sqrt{n})$ ($t\in\Omega(\frac{1}{\theta})$). Choosing $t$ provides $a\approx 1$. 

$Pr_{success}(\cal{A})$ is the probability of measuring a unique basic state $\ket{k}\ket{0}\ket{1}$ among all basic states $\ket{j}\ket{0}\ket{1}$, $j\in\{0, \dots n-1\}$ of the part $\sin((2t+1)\theta)\ket{\psi_{good}}$ of the state $\ket{\psi^t}$ (\ref{psi_t}). From this and from (\ref{psi_0}), (\ref{psi_good}) it follows that
\[
Pr_{cussess}({\cal{A}})=a \frac{1}{p_{good}}\frac{1}{n}= a \frac{1}{\left(\frac{1}{n}\right)\left(1+\sum_{j, j\not= k}\epsilon_j^2\right)n} \ge a \frac{1}{1+(n-1)\epsilon^2} \ge a \frac{1}{1+c(\epsilon)}.
\]

Thus, selecting $t\in\Omega(\sqrt{n})$ and choosing $c>0$ close to 0( chosen according to $\epsilon$ respectively) provides $Pr_{cussess}({\cal{A}}) \approx 1$.

\subsubsection{Estimation of the memory $S({\cal A})$}

The dimension $S({\cal A})$ of the algorithm ${\cal A}$ is defined as the total number of qubits used in the state $\ket{V, \psi_E}$. So we have
 \[
S({\cal A}) = \log n + s +2.
\]

For some selected $c>0$ and $\epsilon  = \sqrt{c/(n-1)}$ we have
\begin{eqnarray}
  S({\cal A}) &=& \log n +\log m + \log{(\epsilon^2(n - 1))} +2  \nonumber \\
&=& \log n +\log m + \log{\epsilon^2} + \log(n-1) + 2 \nonumber \\
&\leq&  2\log{n} + \log m + const  \nonumber
\end{eqnarray}
%
%
\begin{Comment}
    Recall that $\epsilon$ is a parameter of the error-correcting code $E$, and the code $E$ defines the quantum fingerprinting function $\psi_E$. The value of $\epsilon$ plays an important role in determining the pairwise ``almost'' orthogonality ($\epsilon$-orthogonality) of the values of $\psi_{E}$. This is essential for proving the high probability of a correct result from the algorithm.
\end{Comment}

\section{Algorithm  ${\cal A}2$ based on quantum hashing }

In this section, we present a generalization ${\cal A}2$ of the algorithm ${\cal A}$. The generalization is based on the application of quantum functions called ``$\epsilon$-stable''.

\subsection{Quantum $\epsilon$-stable functions}

Let's define an $\epsilon$-stable quantum function following \cite{DBLP:journals/bjmc/AblayevAVZ16} (see also the monograph \cite{aav_book_2023}).

\begin{Definition}\label{e-resistant} \textbf{($\epsilon$-stable function)}
     The quantum function
\[
\psi :\Sigma^m\to ({\mathcal
H}^2)^{\otimes s}, 
\]
that maps words of length $m$ in the finite alphabet $\Sigma$ to a set of $s$-qubit states will be called $\epsilon$-stable, if
for all words $w_j\in\{0,1\}^m$ their quantum images $\ket{\psi(w_k)}$ and $\ket{\psi(w_j)}$ are pairwise $\epsilon$-orthogonal
\[
\left|\langle{\psi(w_k)}|{\psi(w_j)}\rangle\right| \le
\epsilon. 
\]
\end{Definition}
 
\begin{Definition}\label{(m,e,s)-hashfunction} \textbf{(($m$,$\epsilon$,$s$)-quantum hash function)}
The quantum function
\[
\psi :\Sigma^m\to ({\mathcal
H}^2)^{\otimes s}, 
\]
that maps words of length $m$ in the finite alphabet $\Sigma$ to a set of $s$-qubit states will be called ($m$,$\epsilon$,$s$)-quantum hash function if quantum function $\psi$ is $\epsilon$-stable and
\[
s \ll m.
\]
\end{Definition}
The requirement of $\epsilon$-stable property of the $\psi$ function imposes restrictions on the compression ratio. The following lower bound   for $s$ is was presented  in \cite{ablayev2014quantum}.
\begin{Theorem}\label{min_count_qubits}
Let
$\psi : \Sigma^m \rightarrow ({\mathcal
H}^2)^{\otimes s}$  be a  ($m$,$\epsilon$,$s$)-quantum hash function. Then 
\[
s \geq \log{m} - \log{\log{(1 + \sqrt{2/(1 - \epsilon)})}} - 1.
\]
\end{Theorem}
{\em Proof.}  The proof is given in the work \cite{ablayev2014quantum} and the monograph \cite{aav_book_2023}. \Endproof

\paragraph{Transformations $U_\psi $ and  $U^{-1}_\psi$. }

The algorithm ${\cal A}2$ uses the unitary transformation $U_\psi$ on the space $({\mathcal
H}^2)^{\otimes s}$ and its inverse $U^{-1}_\psi$.

The $U_\psi$ transformation is an implementation of the ($m$,$\epsilon$,$s$)-function $\psi$ . For the $U_\psi$ transformation, the following must be true
\[
U_\psi: \ket{0} \xrightarrow{w}  \ket{\psi(w)}. 
\]
Accordingly, for the inverse transformation $U^{-1}_\psi$, the following must be true   
\[
U^{-1}_\psi: \ket{\psi(w)} \xrightarrow{w}  \ket{0}. 
\]

An example of such a function $\psi$ is the fingerprinting function $\psi_E$ based on the error-correcting code $E$. The corresponding transformations $U_{\psi_E}$ and $U^{-1}_{\psi_E}$, which are described in Properties \ref{u_property} and \ref{u-1_property}, are also given.

\subsection{Algorithm ${\cal A}2$}
The algorithm ${\cal A}2$ is a generalization of the quantum algorithm ${\cal A}$ in terms of an arbitrary ($m$,$\epsilon$,$s$)-quantum hash function $\psi$.

 The algorithm ${\cal A}2$ consists of two parts:
\begin{enumerate}
\item First part: preparing the initial state based on sequence $V$.
\item Second part: reading the search word $w$ and searching for its position in the sequence.
\end{enumerate}

Note that the algorithm ${\cal A}2$ has different input data for the two parts: $V$ and the word $w$, respectively.

{\bf Description of the algorithm ${\cal A}2$.}
 \begin{quote}
 \item {\bf Input data:} 
 
{\bf For the first part}: sequence  $V=\{w_0, \dots, w_{n-1}\}$ of binary words with length $m$. 

{\bf For the second part}: Binary word $w$ with length $m$.

\item {\bf Output data:} The index $k$, which denotes the word number $k$ such that $w=w_k$.
\end{quote}

Thus, the algorithm ${\cal A}2$ implements the mapping
\[
{\cal A}2: V,  w \longmapsto k.
\]

\begin{enumerate} 

\item {\bf The first part of the algorithm} consists of preparing the initial state.

According to this sequence $V$, using the transformation $U_\psi$, which defines the ($m$,$\epsilon$, $s$)-quantum hash function $\psi$, a state is prepared:
    \[
\ket{V, \psi} = \frac{1}{\sqrt{n}}\sum\limits_{j=0}^{n-1}\ket{j} \otimes \ket{\psi(w_j)} \otimes \ket{1},
\] 
\item {\bf The second part of the algorithm ${\cal A}2$ -- amplification} 
consists of reading a word to be searched for, $w$, and finding a number $k$ such that $w=w_k$.

\begin{enumerate}
    \item \textbf{Conversion hash transformation -- first level of amplification}
    
    The operator $I^{\otimes \log{n}} \otimes U^{-1}_\psi(w) \otimes I$ is applied to the state $\ket{V, \psi}$,

where $U^{-1}_\psi(w)$ is the inverse transformation controlled by the searched word $w$. We get the state
\[
\ket{V, \psi, w} =
\frac{1}{\sqrt{n}} \sum\limits_{j=0}^{n-1}\ket{j} \otimes  \ket{\phi(w_j, w)} \otimes \ket{1},
\]
where $ \ket{\phi(w_j, w)}= U^{-1}(w)\ket{\psi(w_j)}$ for $j \in\{0,\dots, n-1\}$. 

\item \textbf{Amplification } 

\label{grover_operators} The amplification of basic states $\ket{j}\ket{0}^{\otimes s}\ket{1}$  -- ``good states'' is applied to state $\ket{V, \psi, w}$. The procedure is described in \cite{brassard2002quantum}, and is also discussed in the book \cite{kaye2006introduction}.

\item \textbf{Measurement}

The first $\log{n} +s$ qubits of the final state are measured in a computational basis. If the last $s$ qubits are all zero, then the measurement result $k$ of the first $\log n$ qubits is declared as the index of the word $w_k$ in the sequence $V$, for which $w_k=w$.

\end{enumerate}

\end{enumerate}

\section{Analysis of the algorithm ${\cal A}2$}

Meaningfully, the algorithm ${\cal A}2$ has the following characteristics.

\begin{Theorem} \label{th_aa2_informal}  
For the algorithm ${\cal A}2$, which searches for the occurrence of an element with length $m$ in an unordered sequence of $n$ elements, the following is true
\[ Pr_{success}({\cal A}2) \approx 1, \quad Q({\cal A}2) = O(\sqrt{n}) 
\quad \mbox{ and } \quad
S({\cal A}2)  = \log n + s, 
\]
where $s$ is the number of qubits allocated to the ($m$,$\epsilon$,$s$) - quantum hash function value.
 
\end{Theorem}

\subsection{Proof of the Theorem \ref{th_aa2_informal}.}

The rest (main) two parts of the proof of the statement of the Theorem \ref{th_aa2_informal} a) space complexity of the algorithm and b) the proof of the correctness of the algorithm are completely determined by properties of   $(m,\epsilon,s)$-quantum hash function $\psi$. 

\subsubsection{Estimation of the probability $Pr_{success} ({\cal A}2)$ of the algorithm success and query complexity $Q({\cal A}2)$}
The proof of the correctness of the algorithm  that is, the proof of the high probability $Pr_{success}({\cal{A}}2)$ of the success of the algorithm  follows the proof of   the Theorem \ref{th_aa_formal} with one modification. 

We replace here the Property \ref{buhtman2001} and the  Lemma \ref{m_lemma}  by the  following Property.
\begin{Property}
\label{m_property}
 Let  the function $\psi$ be an $\epsilon$-stable function.
%
%
 Then for the state $ \ket{\phi(w_j, w)}= U^{-1}_{\psi}(w)\ket{\psi(w_j)}$
 \[\ket{\phi(w_j,w )}=\alpha_0\ket{0}+ \alpha_1\ket{1} + \dots +  \alpha_{2^l-1}\ket{2^s-1}  \]
the following is true: If  $w_j=w$, then  
$ \alpha_0=1 
$.
If $w_j\not= w$, then  
   $ \alpha_0 \le \epsilon  
   $.
\end{Property}
The proof of the Property \ref{m_property}  repeats  the proof of the 
Lemma \ref{m_lemma} with the simplification that we do not need to prove the $\epsilon$-orthogonality ($\epsilon$-stable) property of function $\psi$. Remind that such  the $\epsilon$-stability of the function $\psi_E$ is essential point in the proof of Lemma \ref{m_lemma}. But here we have such property ($\epsilon$-stability) from  the definition of ($m$,$\epsilon$,$s$)-quantum hash function  $\psi$.  

Query complexity $Q({\cal{A}}2) = O(\sqrt{n})$ of the algorithm  is determined by the number of applications of Grover iterations to search for occurrences of the word $w$ in a dictionary $V$ consisting of $n$ words. The implementation of the ``query'' part of the algorithm ${\cal A}2$ is described in the section \ref{grover_operators}  and in fact repeats the proof for the presented characteristics in Theorem $\ref{th_aa_formal}$.

We present only the final calculations from the proof.

Let $a=\sin^2((2t+1)\theta)$ is the probability of measuring one of the states $\ket{j}\ket{0}\ket{1}$ after applying all Grover iterations. Note that for small $\theta$: $\sin(\theta)\ge\theta-\delta$ for small $\delta$ ($\sin(\theta)\approx \theta$). In particular, in our case
\[
\sin(\theta)= \sqrt{\frac{1}{n}\left( 1 + \sum_{j=0, j\not= k}^{n-1}\epsilon_j^2\right)}.
\]
We choose $t$ such that $(2t+1)\theta \approx\frac{\pi}{2}$, meaning $t\in\Omega(\sqrt{n})$ ($t\in\Omega(\frac{1}{\theta})$). Choosing $t$ provides $a\approx 1$. 

$Pr_{success}({\cal{A}}2)$ is the probability of measuring a unique basic state $\ket{k}\ket{0}\ket{1}$ among all basic states $\ket{j}\ket{0}\ket{1}$, $j\in\{0, \dots n-1\}$ of the part $\sin((2t+1)\theta)\ket{\psi_{good}}$ of the state $\ket{\psi^t}$ (\ref{psi_t}). From this and from (\ref{psi_0}), (\ref{psi_good}) it follows that
\[
Pr_{cussess}(A)=a \frac{1}{p_{good}}\frac{1}{n}= a \frac{1}{\left(\frac{1}{n}\right)\left(1+\sum_{j, j\not= k}\epsilon_j^2\right)n} \ge a \frac{1}{1+(n-1)\epsilon^2}.
\]
Thus, selecting $t\in\Omega(\sqrt{n})$ and choosing $\epsilon$ close to 0 provides $Pr_{cussess}({\cal{A}}2) \approx 1$.

\begin{Comment} 

Here we would like to comment on the second part of the algorithm (amplification) and its stages as follows.  We  call (following the  \cite{kaye2006introduction}) the basic states in the form $\ket{j}\ket{0}^{\otimes s }\ket{1}$ as ``good states'', which are states whose amplitudes we amplify using well-known quantum amplification procedure. The remaining basic states are called ``bad states''.

Note that there are $n$ good basic states
\[
Good= \left\{ \ket{\bf 0}\ket{0}^{\otimes s}\ket{1}, \dots, \ket{\bf n - 1}\ket{0}^{\otimes s}\ket{1} \right\}
\]
and only (some of them) states $\ket{k}\ket{0}^{\otimes s}\ket{1}$ for which $w_k=w$ among the ``good states'' are ``correct''. 
The necessary distinction between correct and incorrect states among the ``good states'' is implemented by means of the reverse hash transformation. This is the first level of amplification, and it works as follows.

Reverse hashing with a given word $w$ creates a state that is divided into two sets: a set of good states and a set of bad states. The set of good states is further divided into correct states and incorrect states. Reverse hashing operates such that the amplitudes of incorrect states is reduced from the initial value of $1/\sqrt{n}$ to $\epsilon / \sqrt{n}$, while remaining amplitudes remain at $1 / \sqrt {n}$. This is the first step in amplitude amplification. Subsequent $O(\sqrt {n})$ steps of Grover's algorithm amplify the amplitude of both correct and incorrect good states, such that the probability of finding correct states approaches 1, while the probability for incorrect states does not grow significantly.
\end{Comment}

\subsubsection{Estimation of the memory $S({\cal A}2)$}

The space complexity $S({\cal A}2) = \log n + s +1$ is determined by the number of qubits representing the initial state $\ket{V,\psi}$ and its transformations. The essential component of $\log n + s +1$ here is $s$ — the compression ratio achieved by the hash function $\psi$.


Upper bound for $s$ depends on the type of the ($m,\epsilon,s$)-quantum hash function.

\section{Conclusion}
In conclusion, we note that the  algorithm ${\cal{A}}2$ is based on the application of a $(m,\epsilon,s)$-quantum hash function.

We will call a ($m$,$\epsilon$,$s$)-quantum hash function ``good'' if it is highly compressive (that is, the number of qubits allocated for the value of the hash function should be close to the lower bound defined in Theorem \ref{min_count_qubits}). Otherwise, we will consider the ($m$,$\epsilon$,$s$)-quantum hash function to be ``bad''.

Choosing a ``good'' $(m,\epsilon,s)$-quantum hash function allows you to achieve a significant reduction in the number of qubits required for the algorithm to work.

The fingerprinting function described in Section 3, which is based on error-correcting codes, is a ($m$,$\epsilon$,$s$)-quantum hash function. In addition, it is a good hash function:
the estimate of $s$ is defined as $s=O(\log{m})$.

Another function, that is also a ($m,\epsilon,s$)-quantum function, is based on Freivald's finger\-printing. This technique was proposed in the 1970s and has a long history of various ap\-pli\-ca\-tions. In particular, for quantum computations  it has been used to construct efficient quantum automata  \cite{ambainis19981}  and quantum branching programs \cite{ablayev2009algorithms}.

Freivald's finger\-printing also defines a ($m,\epsilon,s$)-quantum function (for the construction, see, e.g., \cite{ablayev2014quantum} and the book \cite{aav_book_2023}).
In the content of this work, such the ($m,\epsilon,s$)-quantum function is also ``good'', i.e. $s=O(\log (cm))$ for $c>1$ and with $\epsilon$ equal to almost $1/c$.

So, the above two ($m$,$\epsilon$,$s$)-quantum hash functions are  ``good''. That is  they use  $s$ qubits which are exponentially less than the length of a word from the dictionary.

An example of a ``bad'' ($m,\epsilon,s$)-quantum hash function is one based on the universal family of linear hash functions \cite{ablayev2014quantum}, see also the book \cite{aav_book_2023}. This hash function is not asymptotically optimal in terms of the number of qubits required for construction.

The study of various ($m,\epsilon,s$)- quantum hash functions, problems of their efficient im\-ple\-men\-tation and development of algorithms based on them continues.

\end{document}